# Stability of hypothetical Ag$^{II}$Cl$_2$ polymorphs under high pressure, revisited – a computational study

Adam Grzelak[1]* and Wojciech Grochala[1]

[1]Center for New Technologies, University of Warsaw, Banacha 2C, 02-097 Warszawa, Poland
*a.grzelak@cent.uw.edu.pl

**Abstract**

A comparative computational study of stability of candidate structures for an as-yet unknown silver dichloride AgCl$_2$ is presented. It is found that all considered candidates have a negative enthalpy of formation, but are unstable towards charge transfer and decomposition into silver(I) chloride and chlorine within the DFT and hybrid-DFT approaches in the entire studied pressure range. Within SCAN approach, several of the "true" Ag$^{II}$Cl$_2$ polymorphs (i.e. containing Ag(II) species) exhibit a region of stability below ca. 20 GPa. However, their stability with respect to aforementioned decomposition decreases with pressure by account of all three DFT methods, which suggests a limited possibility of high-pressure synthesis of AgCl$_2$. Some common patterns in pressure-induced structural transitions observed in the studied systems also emerge, which further testify to an instability of hypothetical AgCl$_2$ towards charge transfer and phase separation.

**Introduction**

Chemistry of silver(II) compounds constitutes a topic of studies that is both demanding – particularly due to the extremely strong oxidizing properties of these compounds – as well as interesting, as evidenced by the body of works discussing them in relation to oxocuprates – a well-known family of precursors for high-pressure superconductors. [1,2] Inspired by experimental works exploring high-pressure phase transitions of AgF$_2$, [3,4] as well as most recent computational study of thermodynamic stability of hypothetical mixed-valence silver fluorides (including at elevated pressure conditions), [5] this work is a continuation to a previous systematic study, which explored relative stability of multiple hypothetical polymorphs of Ag$^{II}$Cl$_2$ – an as-yet unknown analogue of AgF$_2$. [6] The aforementioned study found that a true silver(II) chloride is likely to be unstable towards charge transfer and phase separation into AgCl and Cl$_2$ at ambient pressure conditions. This work aims to extend these considerations into high-pressure regime, in the hope that applying extreme conditions could stabilize Ag$^{II}$Cl$_2$. In particular, the previous study found that a layered, AgF$_2$-type polymorph of AgCl$_2$ could be stabilized at a pressure of ca. 35 GPa, due to relatively low molar volume. [6]

The interest in this particular compound stems from its potential similarity to AgF$_2$, which has recently been shown to be an excellent analog of oxocuprates in terms of structure and very strong magnetic interactions. [2] In fact, Ag$^{II}$Cl$_2$ – if obtained, and providing a suitable structural arrangement – could be expected to host even stronger antiferromagnetic superexchange than AgF$_2$, due to stronger covalence of Ag–Cl bonding, as predicted from differences in electronegativity (Ag: 1.93, Cl: 3.16, F: 3.98 – Pauling scale). Overall, this work is part of a joint computational and experimental effort: to synthesize Ag$^{II}$Cl$_2$ utilizing high pressure and high temperature experimental techniques, coupled with computational methods providing insight into understanding the expected products and phases. On top of that, the added value of studies in high-pressure regime is that the observed changes in structure and bonding induced by pressure can provide meaningful insight into the chemical nature of the studied compounds. [7]

**Results**

Stability of AgCl$_2$ phases

Seven different candidate structures for polymorphs of AgCl$_2$ were considered in this work. Six of them were derived from the previous, ambient-pressure study. [6] They were:

- AgF$_2$ type – corrugated layers made up of [AgCl$_4$] square subunits;
- CuCl$_2$ type – 1D chains made up of [AgCl$_4$] square subunits;
- AuCl$_2$ type – a kind of nanotubular polymorph derived from 0D molecular structure of AuCl$_2$ (see below);
- Ag(I)r type – a structure composed of double layers of rocksalt-type AgCl interspersed with Cl$_2$ bridges;
- Ag(I)h type – similar as above, but with hexagonal double layers of AgCl;
- MnO$_2$ type – a different arrangement of corrugated layers made up of pairs of [AgCl$_4$] square subunits.

Additionally, the high-pressure, nanotubular polymorph of AgF$_2$ (referred to as AgF$_2$ HP) was considered as a candidate. [4] Structures of selected polymorphs described above are presented in fig. 1.

Importantly, Ag(I)r and Ag(I)h polymorphs do not contain Ag(II) species and instead are composed of sub-structures of Ag$^I$Cl and Cl$_2$ molecules. For the remaining five Ag$^{II}$Cl$_2$ polymorphs containing paramagnetic d$^9$ silver cation, magnetic interactions were taken into account:

- AgF$_2$ type: 2D antiferromagnetic (AFM) coupling within layers;
- CuCl$_2$ type: AABB-type AFM coupling within chains, known to exist in CuCl$_2$, [8] and found to be the lowest-energy magnetic solution for AgCl$_2$ in this arrangement; [6]
- AgF$_2$ HP type: magnetic dimers coupled along ~180 degrees Ag-F-Ag bridges within nanotubes;
- MnO$_2$ type: AABB-type AFM coupling – ferromagnetic (FM) between adjacent [AgCl$_4$] squares and AFM between pairs.

The AuCl$_2$ type derives from a structure of gold(I,III) chloride, a mixed-valence compound which consists of Au$_4$Cl$_8$ molecules, each containing two Ag(I) and two Ag(III) species. [9] When this structure is taken as a starting point for geometry optimization, all three computational methods yield a nanotubular polymorph somewhat similar to the AgF$_2$ HP type. However, only the HSE06 approach is able to reproduce the mixed-valence nature of AuCl$_2$ and of the corresponding nanotubular polymorph of AgCl$_2$ derived from the former (as evidenced by two different coordination patterns of Ag sites in that solution). [6] On the other hand, in the PBEsol+U and SCAN calculations, a magnetic model with dimers as in the AgF$_2$ HP type polymorphs was considered.

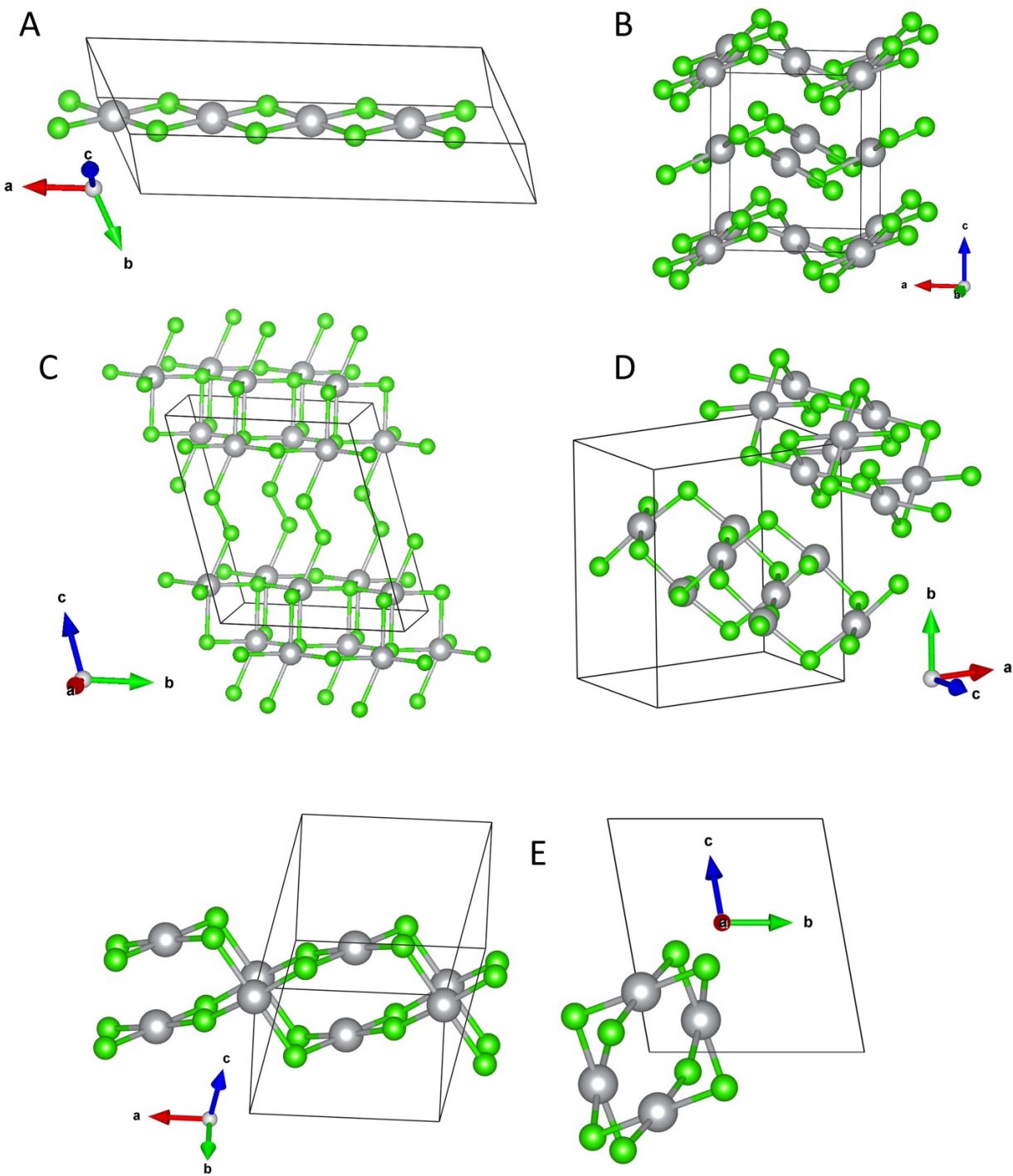

Fig. 1. Selected structures of AgCl$_2$ candidate polymorphs. A – CuCl$_2$-type, B – AgF$_2$-type, C – Ag(I)r, D – AgF$_2$-HP-type, E – AuCl$_2$-type. Ag – grey, Cl – green.

Stability of the studied candidate structures was evaluated using two parameters: (a) **enthalpy of formation** (labelled henceforth as **ΔH$_f$**), according to a reaction:

(Eq. 1) $Ag + Cl_2 \rightarrow AgCl_2$

and (b) **stability towards decomposition** into AgCl and $Cl_2$ (**$\Delta H_r$**), or more precisely, the enthalpy of reaction:

(Eq. 2) $AgCl + \frac{1}{2}Cl_2 \rightarrow AgCl_2$

Defined in this manner, both parameters indicate thermodynamic instability when positive. Fig. 2 shows plots of stability of studied $AgCl_2$ candidate types in terms of $\Delta H_r$ for all three methods used in this work. Enthalpy of Ag + $Cl_2$ mixture relative to AgCl + ½$Cl_2$ is also plotted for comparison.

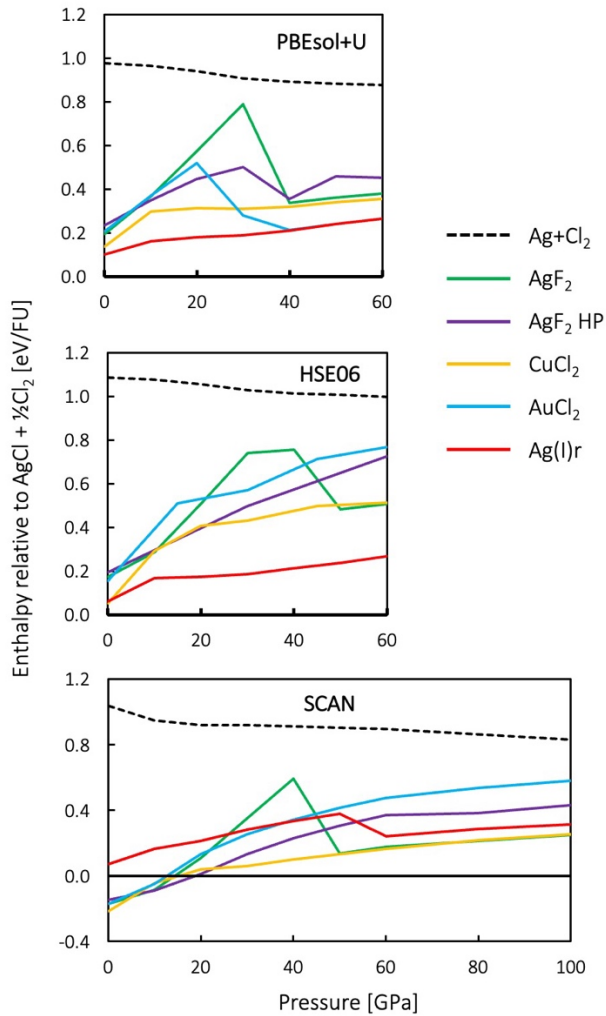

Fig. 2. Stability of selected studied polymorphs of $AgCl_2$, plotted as enthalpy per formula unit relative to AgCl + ½$Cl_2$ ($\Delta H_r$). Top panel – PBEsol+U (GGA functional), middle panel – HSE06 (hybrid functional), bottom panel – SCAN (meta-GGA functional). FU – formula unit.

The studied polymorphs exhibit negative (favorable) enthalpies of formation within the entire studied pressure range in all three computational approaches. This is indirectly visible in fig. 2 as the fact that curves for those polymorphs lie below the curve for Ag + $Cl_2$. On the other hand, they were found to be unstable in terms of $\Delta H_r$ in the entire studied pressure range within PBEsol+U and HSE06. However, results of SCAN calculations indicate moderate stability of $Ag^{II}Cl_2$ polymorphs below ca. 20 GPa. Ag(I)r polymorph is the most stable among $AgCl_2$ candidate structures, according to PBEsol+U and HSE06 results, while this is not the case in SCAN picture. Given that it contains separate sub-phases of AgCl and $Cl_2$, this further points to an instability towards charge transfer and phase separation.

The initial set of calculations was performed within the PBEsol+U approach, as the least computationally demanding. It was found that the Ag(I)h type collapses upon compression to 10 GPa into the Ag(I)r structure and it was not considered any further in the analogous HSE06 and SCAN calculations. Similarly, $MnO_2$ type was also neglected past the PBEsol+U approach, as it was found to be the least stable in terms of $\Delta H_r$ among the studied types. Therefore, these two polymorphs are not taken into account in figures 1, 2 and 3.

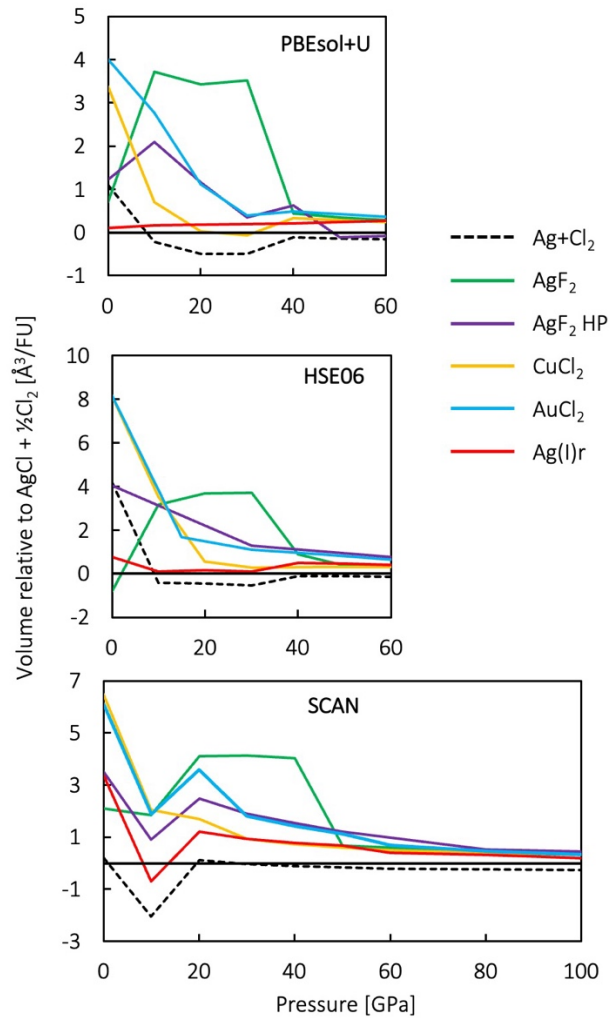

Fig. 3. Volume of selected studied polymorphs of $AgCl_2$ relative to AgCl + ½$Cl_2$. Top panel – PBEsol+U, middle panel – HSE06 (hybrid functional), bottom panel – SCAN (meta-GGA functional). FU – formula unit.

Fig. 3 compares relative volume ($\Delta V_r$) of the studied polymorphs with respect to AgCl + ½$Cl_2$ mixture. Since the $pV$ term becomes a considerable contribution to enthalpy at elevated pressures, the fact that most of the candidates for $AgCl_2$ considered here have a positive $\Delta V_r$ can be seen as an important factor leading to their relative instability. However, it should be noted that only HSE06 accurately reproduces ambient-pressure volumes of AgCl and $Cl_2$. PBEsol+U underestimates the volume of AgCl and $Cl_2$, while SCAN – that of $Cl_2$. Therefore, $\Delta V_r$ values at 0 GPa in fig. 3, and by extension – initial compressibilities – should be taken with a grain of salt. A noticeable dip at 10 GPa within the SCAN approach is likely a manifestation of this. The computationally demanding HSE06 results are likely to be the most correct.

Some of the features in figures 2 and 3, such as abrupt changes of relative energies or volumes, and apparent convergence of plots corresponding to different polymorphs are indicative of structural transitions. The nature and implication of those transition will be discussed in the next section.

Pressure-induced structural transitions

As an introduction to analysis of structural transitions of $AgCl_2$ polymorphs, let us first discuss the Ag(I)r solution. As mentioned before, this structure is made up of subunits of rocksalt-type AgCl and of $Cl_2$ molecules. Within the studied pressure range, it undergoes structural rearrangements, which can be approximated as a sequence of deformations of the AgCl double layers, leading from a fundamentally NaCl-like coordination patterns to CsCl-like patterns, with increasing coordination number of Ag atoms. Importantly, it should be stressed again that this structure emerged as one of the lowest-energy solutions in an evolutionary algorithm structural search reported in the previous contribution. [6] It should not be treated as a viable candidate for the structure of $AgCl_2$, but rather as a manifestation of the proclivity of the studied system towards charge transfer and phase separation. The case for instability of $Ag^{II}Cl_2$ towards these processes is further strengthened by the fact that the Ag(I)r polymorph remains the most stable (with respect to $\Delta H_r$) throughout the studied pressure range in both PBEsol+U and HSE06 approaches. A noticeable drop in $\Delta H_r$ between 50 and 60 GPa for this polymorph in the SCAN approach was the reason for extending the studied pressure range to 100 GPa in this case. However, this drop is a result for NaCl-CsCl-like transition in the AgCl subphase, which is more abrupt than in analogous PBEsol+U and HSE06 calculations. No further phase transitions for Ag(I)r polymorph are observed up to 100 GPa.

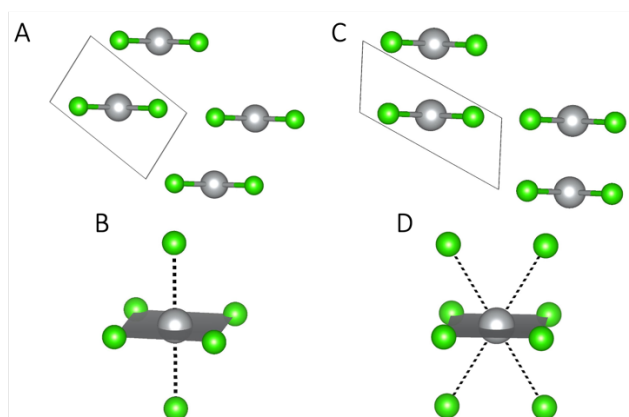

Fig. 4. Structural changes in $CuCl_2$-polymorph. A and C – view along the direction of chain propagation at 10 and 20 GPa, respectively. B and D – view of local coordination of Ag at 10 and 20 GPa, respectively. Ag – grey, Cl – green.

The $CuCl_2$-type polymorph emerges as the most structurally robust in this study, as it does not undergo any collapse or drastic deformation in the studied pressure range, maintaining a relatively low $\Delta V_r$ in all three methods. The only change to the structure of this polymorph occurs in terms of arrangement of chains relative to one another. Up to 10 GPa (in all three computational approaches), the chains are positioned as in fig. 4A, resulting in octahedral coordination of Ag atoms (fig. 4B). The octahedra are elongated by 29%, 29% and 25%, according to PBEsol, HSE06 and SCAN, respectively. By account of all three methods, the arrangement changes between 10 and 20 GPa, leading to a 4+4 coordination of Ag atoms, with the 4 inter-chain contacts longer by 23 to 33%, depending on the method (fig. 4D). This is achieved in different ways: in PBEsol+U and HSE06 results, the chains move relative to one another in a direction perpendicular to direction of propagation (fig. 4C). In SCAN approach, this is achieved through a change of one of the unit cell angles, which results in sliding the chains relative to each other. The resulting 4+4 coordination is the same in all cases (fig. 4D), but the longer Cl contacts are aligned parallel (in SCAN) or perpendicular (in PBEsol+U and HSE06) to direction of propagation. Further compression to 30 GPa transforms the structure into that seen at 20 GPa in PBEsol+U and HSE06 pictures. All of these transitions can be seen as a means to achieve a more efficient packing of chains (as evidenced by increasing coordination number of Ag). Changes in local coordination of Ag atoms in this polymorph are plotted in the top panel of fig. 6.

Another result of these rearrangements is a reduction of Cl…Cl distances between neighboring chains. These distances drop from ca. 3.8 to 2.7 Å between 0 and 60 GPa in HSE06, compared to 3.2 to 2.5 Å in PBEsol+U and 3.5 to 2.8 Å in SCAN. Recall that only HSE06 correctly reproduced the ambient pressure (low-temperature) volume of solid molecular $Cl_2$ at 0 GPa, so these results further testify to the superiority of HSE06 in describing weak interactions compared to the other two methods utilized here (and free from the explicit van der Waals terms). Importantly, all of these distances are larger than the Cl-Cl bond in solid molecular $Cl_2$, which remains at ca. 2.00 Å and contracts very little (less than 0.05 Å) with compression in results from all three methods. Additionally, this polymorph was further optimized at 100 GPa with PBEsol+U and SCAN methods and does not undergo any structural modifications in that pressure range.

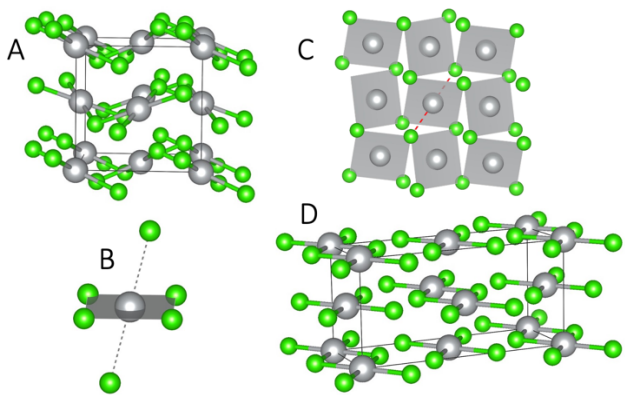

Fig. 5. Structural features of $AgF_2$-type polymorph: A – ambient-pressure structure; B – local coordination of Ag in the ambient-pressure structure; C – one layer at 40 GPa in HSE06 approach. Note that axial Cl atoms are now within the same layer, a pair of those contacts is marked with red dashed line; D – structure resulting from phase transition between 40 and 50 GPa (30 and 40 GPa in PBEsol+U approach). Note the $CuCl_2$-like chains. Ag – grey, Cl – green.

$AgF_2$-type polymorph undergoes substantial structural rearrangements with increasing pressure. At 0 GPa, it adopts a layered structure (fig. 5A), in which every Ag atom forms 4 in-layer bonds with Cl atoms, with two additional Cl atoms from adjacent layers together constituting a distorted octahedral coordination (fig. 5B). The axial Cl contacts are noticeably longer (by 23%, 33% and 27% in PBEsol+U, HSE06 and SCAN, respectively) than equatorial ones. The same phenomenon is observed in $AgF_2$ and in general, elongated octahedral coordination is a well-documented phenomenon in silver(II) fluorides (which includes ternary compounds). [10] This is usually attributed to Jahn-Teller effect, whereby a vibronic instability leads to elongation or contraction of bonds along one of the three axes of octahedron, which lowers the overall electronic energy. Upon compression, this elongation is reduced in AgCl$_2$ down to 1%-2% at 30 GPa in both PBEsol+U and SCAN results. Further increase of pressure (40 GPa in PBEsol+U and 50 GPa in SCAN) leads to a structural transition into a polymorph consisting of 1D chains similar to those found in CuCl$_2$-type polymorph (fig. 5D). In the HSE06 picture, the contraction is less pronounced – down to 10% at 30 GPa, and the transition to the chain polymorph occurs via a different layered structure observed at 40 GPa, where the layers become more corrugated and more separated from each other (fig. 5C). Ag atoms retain an approximately octahedral coordination, but the axial Cl contacts are now within the same layer as [AgCl$_4$] squares. The elongation of octahedra due to Jahn-Teller effect is still noticeable (9%). Further compression to 50 GPa leads to a collapse to chains as in the other two methods. Changes in local coordination of Ag atoms in this polymorph are plotted in the bottom panel of fig. 6.

The transitions described above can be seen as an abrupt drop in relative enthalpy in fig. 2. In the previous contribution discussing relative stability of AgCl$_2$ candidate structures, AgF$_2$-type emerged as the most likely candidate at higher pressures due to its comparatively low molar volume among the considered structures. [6] However, it appears that compression of AgF$_2$-type produces a lot of strain in the structure, as evidenced by a strong increase of ΔH$_r$ (fig. 2), which is released through the aforementioned transition.

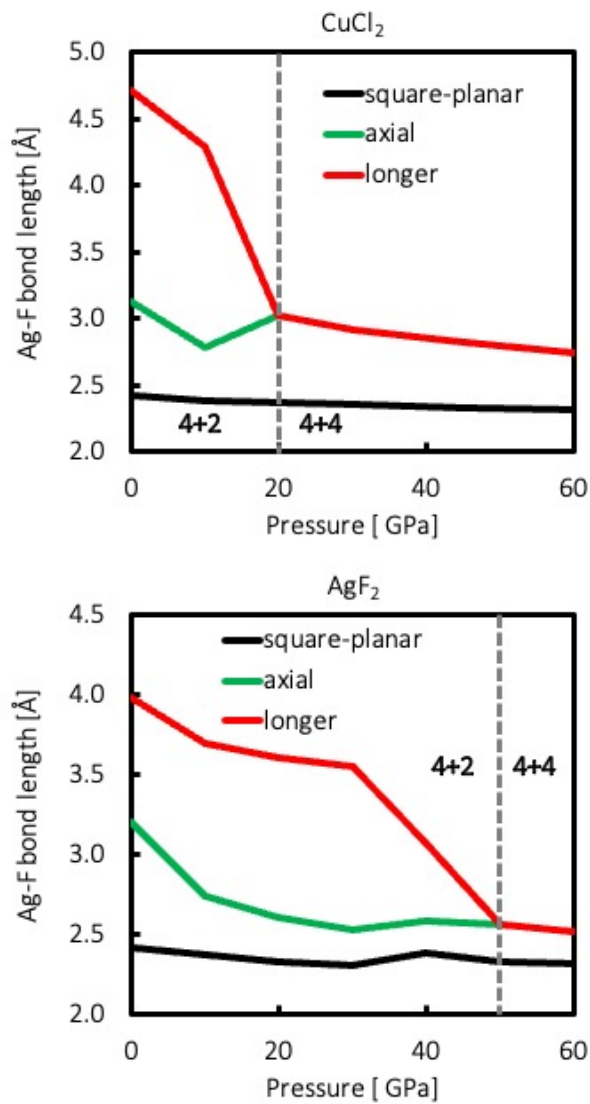

Fig. 6. Pressure dependence of Ag-F distances in CuCl$_2$-type and AgF$_2$-type polymorphs in the HSE06 picture.

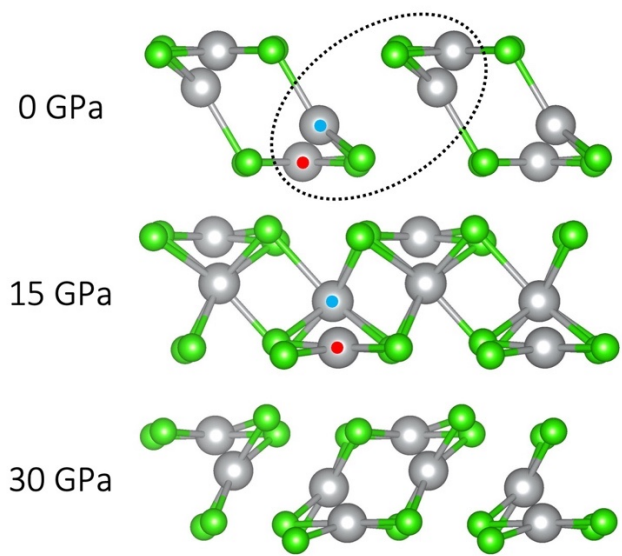

Fig. 7. Structural transition of AuCl$_2$-type polymorph in the HSE06 picture. Nanotubes are viewed along the axis of propagation. Blue and red circles indicate Ag(I) and Ag(III) species, respectively.

Arguably the most interesting pattern of pressure-induced transitions can be observed in the AuCl$_2$-type nanotubular polymorph. The differences in outcomes of compression between the three computational approaches are the most pronounced for this system, although upon closer look we can identify their fundamental similarity. Recall that in the HSE06 approach, AuCl$_2$-type is mixed-valent: Ag(I) species are connected to 3 Cl atoms in an approximately flat triangular pattern, while the Ag(III) sites appear as [AgCl$_4$] square units, which are analogous to those in AgF$_3$. [11] At 0 GPa, the triangular contacts average 2.53 Å, while the bonds in square units are 2.29 Å, which is even shorter than for Ag(II) in [AgCl$_4$] square in CuCl$_2$-type and AgF$_2$-type polymorphs at the same pressure. This supports the assignment of the sites as Ag(I) and as Ag(III), respectively. Compression to 15 GPa induces a change in local coordination of the Ag(I) species, which picks up 4 Cl atoms along the axis perpendicular to the plane of the former triangle, resulting in a 4+3 coordination, with an average bond length of 2.61 Å. Meanwhile, the Ag(III) subunit retains a square coordination with a shorter average bond length of 2.27 Å. Further compression to 30 GPa leads to a rearrangement of nanotubes, which are now formed by a different combination of Ag and Cl atoms; importantly, all Ag atoms are coordinated by 4 Cl atoms in an approximately square-planar manner, with an average for the formerly Ag(III) sites at 2.33 Å and the formerly Ag(I) ones – 2.38 Å. This convergence of the two sites in terms of local coordination points to a comproportionation process, whereby all Ag sites are now nominally Ag(II) species. The transition described above is shown in fig. 7.

A very different scenario is observed in the SCAN picture (fig. 8). At 0 GPa, the two Ag sites are already equivalent, but upon compression to 10 GPa, the local square-planar coordination of one of them is rotated by 90 degrees i.e., two of its four nearest Cl neighbors are substituted for another two, which leads to connections between nanotubes in the *c* direction. Further compression to 30 GPa results in an inward contraction of individual nanotubes. The local coordination of the Ag sites marked with red circle changes to more uniformly octahedral – a fifth Cl atom is picked up from a neighboring nanotube along the axis perpendicular to the [AgCl$_4$] plane. The average length of the five Ag–Cl bonds for this Ag site is 2.51 Å

(2.45-2.55 Å). The sixth nearest neighbor, on the opposite side of the former [AgCl$_4$] is actually another Ag atom, located at a distance of 2.70 Å. (This connection is marked with a dashed line in fig. 8.) Importantly, this new Ag…Ag contact is consistently shorter than the Ag–Ag distance in metallic silver at corresponding pressures (from SCAN calculations): 2.70 Å vs. 2.72 Å (30 GPa), 2.64 Å vs. 2.69 Å (40 GPa) and 2.60 Å vs. 2.67 Å (50 GPa). This leads us to infer a formation of a weak Ag-Ag interacion in this polymorph, which is a very interesting finding and reminiscent of a recent work, where a silver subchloride Ag$_8$Cl$_6$ was predicted, featuring [Ag$_6$] subunits within its structure. [12] Further compression above 50 GPa leads to a substantial structural rearrangement: AgCl-like chains with square cross-section are formed, interspersed with Cl$_2$ molecules.

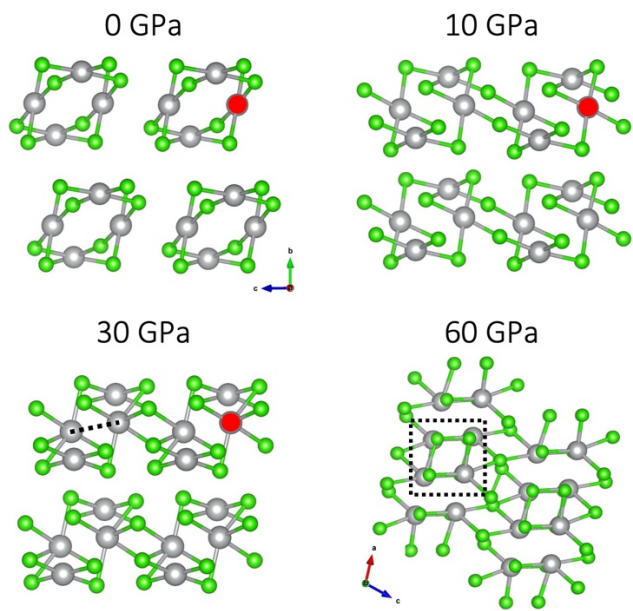

Fig. 8. Structural transition of AuCl$_2$-type polymorph in the SCAN picture. Nanotubes are viewed along the axis of propagation from 0 to 30 GPa. Note a different projection at 60 GPa, along the newly formed [AgCl] square nanowires. Red circles mark the Ag atom which experiences the pronounced changes in local coordination in the 0-30 GPa range.

A similar picture emerges from the PBEsol+U approach. As in SCAN results, compression to 10 GPa leads to a rearrangement of local coordination of half of Ag sites. The subsequent contraction of nanotubes and formation of short Ag…Ag contacts is observed at a lower pressure of 20 GPa (compared to 30 GPa in SCAN). Upon further pressure increase to 30 GPa, a structure similar to Ag(I)r polymorph is formed, consisting of double layers of rocksalt-like AgCl interspersed with Cl atoms. The difference is that here, no discernible Cl–Cl molecules are formed – the distance between Cl atoms lying between AgCl layers is ca. 2.3 Å, compared to ca. 2.0 Å in solid molecular Cl$_2$ and in Ag(I)r polymorph. This is likely an artificial result, which will be further discussed in the "Electronic structure" section. Finally, compression to 40 GPa leads to rearrangement within AgCl layers, which increases the coordination number of Ag from 6 to 7 and the coordination environment resembles that in CsCl. Cl$_2$ molecules between the AgCl layers, characteristic of Ag(I)r polymorph, can also be discerned. Indeed, as can be seen in fig. 2, this solution also converges with Ag(I)r polymorph in terms of ΔH$_r$.

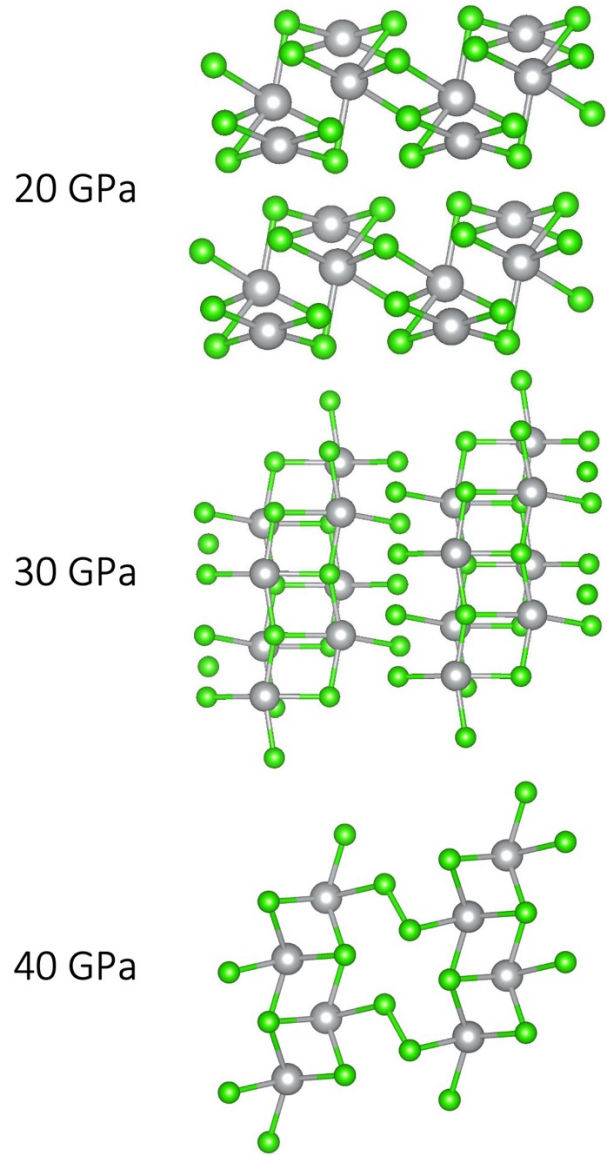

Fig. 9. Structural transitions of AuCl$_2$-type polymorph in the PBEsol+U picture.

As in the case of Ag(I)r solutions, formation of separate domains of AgCl and Cl$_2$ during structural transitions of AuCl$_2$-type in PBEsol+U and SCAN picture, can be interpreted as another manifestation of the system's tendency for AgCl + ½Cl$_2$ phase separation, rather than as viable structural candidates. However, it should be pointed out that sodium and potassium chlorides with exotic stoichiometries as e.g. NaCl or Na$_3$Cl, have been predicted in the past. [13,14]

Nanotubular AgF$_2$-HP-type polymorph undergoes a structural collapse above 30 GPa in PBEsol+U results and above 60 GPa in SCAN results. The final structures do not resemble any of those discussed above; rather, they feature domains of connections between Ag and Cl atoms which do form any extended and discernible pattern, and are instead interspersed with Cl$_2$ molecules. Since those solutions are consistently very high in relative enthalpy (ΔH$_r$), they will not be further analyzed here.

Electronic properties

As inferred from the previous paper, [6] antiferromagnetic superexchange in Ag$^{II}$Cl$_2$ can, in principle, be expected to be strong, since Ag–Cl bonding in this hypothetical compound would likely be more covalent in nature than in its AgF$_2$ counterpart. Of course, AgCl – the only currently known binary combination of silver and chlorine – is an ionic solid, as is AgF. However, previous studies of AgF$_2$ demonstrated a covalent character of Ag–F bonding in that compound, evidenced by X-ray photoelectron spectroscopy [15] and by optical spectra. [16] In the former study, covalency increased in the sequence AgF → AgF$_2$ → AgF$_3$. Therefore, it is reasonable to expect a similar trend in AgCl$_x$ compounds. On the other hand, Cl$^-$ anions are larger and more diffuse and are therefore softer Lewis bases than F$^-$ anions, which makes them more vulnerable to the strongly oxidizing properties of Ag(II) cations. In addition, increasing pressure and the consequent reduction of interatomic distances increases orbital overlap, leading to broadening of electronic bands, which ultimately results in metallization of most known compounds (both ionic and covalent). [7]

Electronic properties of the studied AgCl$_2$ candidate polymorphs were scrutinized in terms of (a) magnetic moments on Ag atoms and (b) fundamental band gap at the Fermi level in electronic density of states (eDOS) graphs. As it turns out, changes in the two parameters are strongly correlated in that the pressure at which the band gap closes coincides with disappearance of magnetic moment on Ag atoms.

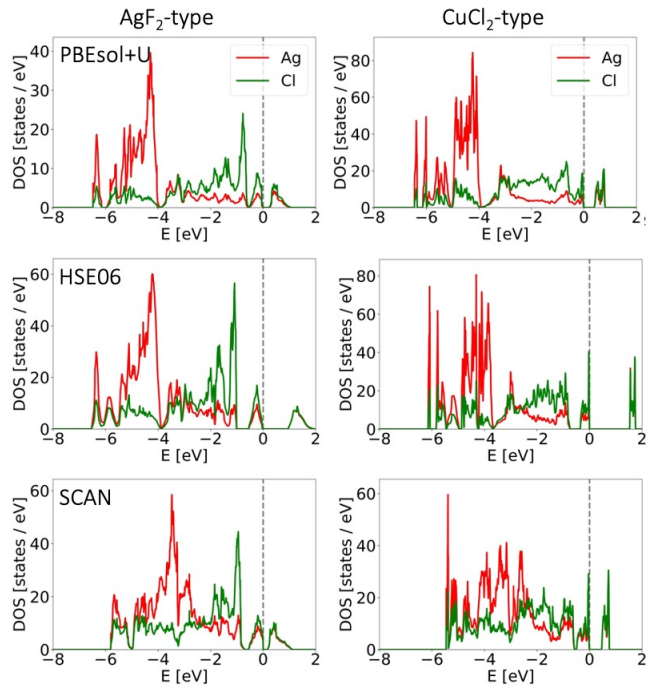

Fig. 10. Comparison of eDOS plots for AgF$_2$-type and CuCl$_2$-type between different computational methods at 0 GPa.

An example of eDOS plots – for AgF$_2$-type and CuCl$_2$-type – at 0 GPa and comparison between the three computational methods is presented in fig. 10. One noticeable feature is the composition of conduction band. In principle, AgCl$_2$, just like the known AgF$_2$, is expected to be a charge-transfer insulator, where the band gap arises between filled nonmetal states and empty metal states (upper Hubbard band, UHB). [17] While this is certainly the case in AgCl$_2$, we can see a substantial admixing of Cl states to the conduction band, with an almost perfect overlap and approximately equal contributions from Ag and Cl states. This indicates a strong covalence of the Ag–Cl bonds that is comparable or indeed even stronger than in $AgF_2$. [15] It should also be pointed out that "insulator" in this case refers to a non-zero band gap resulting from electronic correlation as per the aforementioned Zaanen-Sawatzky-Allen model. [17] Clearly, with a band gap in the range 0.2-1.5 eV (depending on the method) (fig. 10), the two $AgCl_2$ polymorphs in question can be more accurately described as semiconductors.

Table 1 compares metallization pressure for selected polymorphs and between three methods. Metallization pressure is defined here as the lowest pressure point at which a solution exhibits null magnetic moments and null band gap. It can be seen that this value is the highest in HSE06 results. This coincides with the observation that the band gap is the largest for $AgF_2$-type and $CuCl_2$-type polymorphs in the HSE06 picture (fig. 10). The differences between methods likely stem from the inclusion of exchange correlation in the hybrid-DFT-type functional, while in GGA-type PBEsol+U approach, electronic correlation, which is crucial for modelling open-subshell systems such as Ag(II) compounds, is only taken into account through *U* and *J* parameters (mentioned in "Methods" section). HSE06 results can likely be considered the most accurate in the case of this system.

Table 1. Pressure of metallization for different polymorphs of $AgCl_2$.

|  | PBEsol+U | HSE06 | SCAN |
|---|---|---|---|
| $AgF_2$-type | 20 GPa | >60 GPa* | 10 GPa |
| $CuCl_2$-type | 10 GPa | 45 GPa | 10 GPa |
| $AgF_2$ HP-type | 0 GPa | 60 GPa | 10 GPa |
| $AuCl_2$-type† | 0 GPa | 15 GPa‡ | 10 GPa |

Footnotes: *Although transformed into chains similar to $CuCl_2$-type, the polymorph retains residual magnetic moment and non-zero band gap at the maximal studied pressure of 60 GPa. Magnetic coupling is FM intra-chain and AFM inter-chain; †Reopening of band gap occurs at higher pressures – see text. ‡This polymorph is mixed-valent in HSE06 picture and is non-magnetic at 0 GPa.

The phase-separated Ag(I)r polymorph remains insulating within the studied pressure range and by account of all three methods. Structural transitions into phase-separated solutions observed for $AuCl_2$-type polymorph, which were discussed in the previous section, are also associated with reopening of the band gap. That is because constituent parts of these solutions – sublattices made up of ionic AgCl and of molecular $Cl_2$ – are insulators. The fact that the solution for $AuCl_2$-type polymorph at 30 GPa in PBEsol+U picture retains metallic character after the afomentioned transition stems from the presence of dangling, unpaired Cl atoms on the outside of AgCl layers. Such arrangement should be unstable towards a Peierls distortion and formation of $Cl_2$ molecules, which is indeed what happens upon further compression to 40 GPa, and in SCAN picture. This testifies to the relatively poor suitability of PBEsol+U for describing electronic correlation.

**Discussion**

The results presented above shed some light on properties and prospects of synthesis of the hypothetical $AgCl_2$. Structural transitions of $CuCl_2$ and $AgF_2$ types reveal the tendency of the system to avoid repulsion between axial Cl atoms and filled $d(z^2)$ orbital of Ag atoms – through relative displacement of chains in the former and through transition from a layered 2D structure into 1D chains in the latter. A similar tendency is seen in pressure-induced phase transitions of $AgF_2$, where the high-pressure nanotubular structure can be viewed as a means both to maximize coordination number and to minimize the repulsion from the Ag d($z^2$) lone pair. [4] The transitions of AuCl$_2$-type and the relative stability of Ag(I)r polymorph firmly indicate that Ag$^{II}$Cl$_2$ is unstable towards charge transfer and phase separation into AgCl and Cl$_2$.

It is important to note that the overall picture which emerges from data presented here is remarkably consistent across the three computational methods employed: GGA DFT (PBEsol+U), hybrid-DFT (HSE06) and meta-GGA DFT (SCAN). The observed structural and electronic changes are essentially the same between the three computational methods, differing only in terms of pressure at which they occur. Fundamentally, phase transitions observed in the studied polymorphs unfold upon pressure-induced decrease in distance between 1D or 2D structural constituents (chains, nanotubes, layers), which entails overcoming weak repulsive interactions between them, and the HSE06 functional, among the three methods utilized here, is best suited for description of those interactions. In particular, this work shows that the prediction of stability of AgF$_2$-type polymorph, which was based on a reasonable extrapolation from ambient-pressure data, nevertheless turned out not to be false. [6]

Of course, this work does not exhaust the list of possible candidates for the structure of AgCl$_2$. Although the studied candidates generally retain a positive enthalpy with respect to decomposition into AgCl+½Cl$_2$, it is worth noting that the CuCl$_2$-type chain structure does not collapse into phase-separated polymorph even at 100 GPa in PBEsol+U and SCAN results. Analogous HSE06 calculations at 100 GPa were not performed, but based on the observation made here that transition pressures are reliably the highest within HSE06 approach, it is reasonable to predict that such collapse would not be seen in HSE06 picture at 100 GPa, either. The apparent lack of such transition pathway could mean that, in principle, obtaining it as a metastable phase could be possible if e.g., elevated temperatures are used together with moderate pressures. Our previous results [6], as well as earlier predictions by other authors [18,19], all consistently suggest that Ag$^{II}$Cl$_2$, if ever obtained, would likely be metastable.

Although the current study was focused on AgCl$_2$ stoichiometry only, the results of this study permit us to extrapolate observed trends towards the AgCl$_3$. The tendency of AgCl$_2$ stoichiometry to undergo decomposition to (AgCl)(Cl$_2$)½ may suggest that even at larger Cl contents, i.e. for AgCl$_3$ stoichiometry, one will observe phase separation to (AgCl)(Cl$_2$), or Ag$^+$(Cl$_3^-$). [20] A similar result is indicated by theoretical study for isolated AgCl$_3$ molecules in the gas phase. [21]

**Methods**

Calculations were carried out using VASP software. [22–26] Overall, three different computational methods were utilized (underlined are the names by which they are referred to throughout this work):

I. <u>PBEsol+U approach.</u> GGA-type Perdew-Burke-Ernzerhof functional adapted for solids (PBEsol) [27] was used, additionally taking into account Coulombic interactions between d electrons through U and J parameters [28] explicitly set to 5 eV and 1 eV, respectively, and with correction for van der Waals interactions. [29] Plane-wave cutoff energy was set to 800 eV and k-space sampling of ca. 2π x 0.04 Å$^{-1}$ was used, with self-consistent-field convergence criterion of 10$^{-7}$ eV. A denser k-spacing of ca. 2π x 0.03 Å$^{-1}$ was used for electronic density of states (eDOS) calculations.

II. <u>HSE06 approach.</u> Hybrid-DFT HSE06 functional was utilized. [30] Due to higher computational load of this method, a coarser k-space sampling of ca. 2π x 0.04 Å$^{-1}$ was used and plane-wave cut-off was set to 520 eV, with self-consistent-field convergence criterion of 10$^{-7}$ eV.

III. <u>SCAN approach</u>. Meta-GGA-type, strongly constrained and appropriately normed (SCAN) functional was used, [31] with correction for van der Waals interactions. [32] Plane-wave cut-off energy was set to 800 eV and k-space sampling of ca. 2π x 0.035 Å$^{-1}$ was used, with self-consistent-field convergence criterion of 5 x 10$^{-7}$ eV. A denser k-spacing of ca. 2π x 0.025 Å$^{-1}$ was used for eDOS calculations.

The studied pressure range was 0 to 60 gigapascals (GPa), additionally extended to 100 GPa for SCAN calculations. Pressure step was 10 GPa up to 60 GPa (or in some cases 15 GPa in HSE06 approach) and 20 GPa above 60 GPa. $Cl_2$ was considered in its solid polymorph with *Cmca* space group, which is known to be stable in the entire pressure range considered here. [33] Similarly, metallic silver is also stable in its fcc structure in that range. [34] Known phase transitions of AgCl were taken into account when calculating relative enthalpy of $AgCl_2$ polymorphs. [35] A primitive unit cell of the KOH-type, high-pressure polymorph of AgCl was utilized in calculations, since it can be used to accurately describe the continuous nature of NaCl-KOH-TlI-CsCl sequence of phase transitions of AgCl. [35,36]

**Acknowledgments**

A.G. would like to thank Dr. Mariana Derzsi for fruitful discussions and comments at the initial stage of this work's development. A.G. acknowledges the contribution from the Polish National Science Center (NCN) – Preludium project no. 2017/25/N/ST5/01976. This research was carried out with the support of the ICM computer center, University of Warsaw under grant SAPPHIRE (no. GA83-34 and G85-892).

**Author contributions statement**

A.G. and W.G. conceived the study. A.G. performed computations, analyzed the results, made the figures and wrote the manuscript. W.G. reviewed the manuscript.